\documentclass[prd,altaffilletter,twocolumn,superscriptaddress,floatfix,nofootinbib]{revtex4-2} 
\usepackage{amsfonts,amsmath,graphicx,bm}

\usepackage{tikz-feynman}
\usepackage{ifpdf}
\usepackage{multirow}
\usepackage[colorlinks=true, linkcolor=red, urlcolor=red, citecolor=red]{hyperref}
\usepackage{url}
\usepackage{todonotes}
\usepackage{float}
\usepackage{graphicx}
\usepackage{mathtools}
\usepackage[yyyymmdd,hhmmss]{datetime}
\usepackage{color}
\usepackage{bbold}

\def\today{\number\day\space\ifcase\month\or
January\or February\or March\or April\or May\or June\or
July\or August\or September\or October\or November\or December\fi
\space\number\year}

\usepackage{footmisc}

\usepackage{braket}
\usepackage{slashed}

\begin{document}
\title{Precise prediction of the decay rate for $\eta_b\to \gamma \gamma$  from lattice QCD}


\author{Brian~\surname{Colquhoun}} 
\email[]{Brian.Colquhoun@glasgow.ac.uk}
\affiliation{SUPA, School of Physics and Astronomy, University of Glasgow, Glasgow, G12 8QQ, UK}

\author{Christine~T.~H.~\surname{Davies}} 
\email[]{Christine.Davies@glasgow.ac.uk}
\affiliation{SUPA, School of Physics and Astronomy, University of Glasgow, Glasgow, G12 8QQ, UK}

\author{G.~Peter~\surname{Lepage}} 
\affiliation{Laboratory of Elementary Particle Physics, Cornell University, Ithaca, New York 14853, USA}

\collaboration{HPQCD Collaboration}
\homepage[URL: ]{https://www.physics.gla.ac.uk/hpqcd/}


\newcommand{\finalRb}{0.467(11)}
\newcommand{\finalFb}{0.01754(50)\, \mathrm{GeV}^{-1}}
\newcommand{\finalWidthEtabGammaGamma}{\Gamma (\eta_b \to \gamma \gamma) = 0.559(32)_{\text{fit}}(1)_{\text{syst}} \: \mathrm{keV}}
\newcommand{\finalTotalWidthEtab}{ 13.81(^{+17}_{-28})_{\text{NRQCD}}(79)_{\text{LQCD}} \: \mathrm{MeV}}
\newcommand{\finalupsratio}{2.31(13)_{\eta_b}(7)_{\Upsilon}}
\newcommand{\upsexptratio}{2.40(14)_{\eta_b}(3)_{\Upsilon}}


\begin{abstract}
\noindent We calculate the decay rate for $\eta_b \to \gamma \gamma$  in lattice QCD for the first time, providing a precise prediction for the Belle II experiment. Our calculation includes $u$, $d$, $s$ and $c$ quarks in the sea, using gluon field configurations generated by the MILC collaboration, at three values of the lattice spacing from 0.06 fm to 0.03 fm. All quarks are treated in the Highly Improved Staggered Quark formalism, which enables us to reach the $b$ quark mass for our valence quarks on these fine lattices. We calculate quark-line connected correlation functions only. By working at additional heavy quark masses between those of $c$ and $b$ we map out the behaviour of the ratio $f_{\eta_h}/(M_{\eta_h}^2F_{\eta_h}(0,0))$, where $f$ is the decay constant, $M$, the mass and $F(0,0)$, the form factor for decay to two on-shell photons for the pseudoscalar heavyonium meson, $\eta_h$. This ratio takes the approximate value 0.5 in leading-order nonrelativistic QCD (NRQCD) but we are able to give a much more accurate analysis than this. 
Focussing on the $b$ quark mass, we find a ratio of $\finalRb$, giving $\finalWidthEtabGammaGamma$. Combined with a value for the branching fraction from NRQCD, our result can be used to determine the total width of the $\eta_b$ with a 6\% uncertainty. 
\end{abstract}

\maketitle


\section{Introduction}
\label{sec:intro}
The annihilation of a pseudoscalar meson  to two photons can provide a very `clean' test of the Standard Model, since it is a purely electromagnetic process. The rate depends on the strong interaction effects that dictate the internal structure of the meson and hence the form factor for the decay, so a comparison of theory and experiment can provide a stringent test of our understanding of QCD. There are challenges to studying this decay mode, however. 

Recently the theoretical picture for $\eta_c\to\gamma\gamma$ decay was transformed with the first lattice QCD calculation with realistic sea quark content~\cite{Colquhoun:2023zbc}. The result, with a total uncertainty of less than 1\%, is much more accurate than that possible with earlier non-lattice methods. The experimental picture is less clear, however, with the Particle Data Group global fit of $\eta_c$ decay rates~\cite{ParticleDataGroup:2024cfk} having a $\chi^2$ of 184.6 for 94 degrees of freedom.  There are several consistent results from different experiments for the process $\gamma\gamma\rightarrow \eta_c\to K\overline{K}\pi$ and the average value for $\Gamma(\eta_c \to \gamma\gamma)$ inferred from  these is in reasonable agreement with the lattice QCD result, but with a much larger (10\%) uncertainty. Improved experimental results will be needed to match the accuracy of lattice QCD. 

Here we will study $\eta_b \to \gamma\gamma$. There are no experimental signals for this decay mode as yet~\cite{ParticleDataGroup:2024cfk,ALEPH:2002xdz}, but it should be accessible to Belle II~\cite{Belletalk}, so a prediction of the rate now is very timely.  
Theoretical calculations to date have relied on nonrelativistic approaches and are limited by systematic uncertainties associated with missing relativistic and radiative corrections. Lattice QCD calculations, on the other hand, promise the possibility of a result free of these uncertainties. Instead there are statistical errors and systematic errors from the extrapolation to the zero lattice spacing, continuum limit. The continuum extrapolation is particularly difficult for $b$ quarks because they have large mass. Very good control of discretisation errors is required, such as that possible using the Highly Improved Staggered Quark (HISQ) formalism~\cite{Follana:2006rc}. In this formalism tree-level discretisation errors appear first at $\mathcal{O}(a^4)$ (where $a$ is the lattice spacing) since tree-level $\mathcal{O}(a^2)$ errors are removed. For heavy-quark physics we fix the `Naik' coefficient of the improved derivative in the HISQ action following~\cite{Follana:2006rc,Monahan:2012dq}. 
An analysis of the HISQ action at large quark mass~\cite{Follana:2006rc,Colquhoun:2025pnh} then shows that discretisation errors only affect subleading $\mathcal{O}(v^4)$ terms in a nonrelativistic expansion in powers of the quark velocity, $v$, appropriate to the heavyonium physics we consider here.  

The hadronic quantity that parameterises the 2-photon decay rate is the form factor $F(0,0)$, where the two zeroes denote that both photons are on-shell. 
The decay rate is related to the form factor for a pseudoscalar heavyonium meson $\eta_h$ by:
\begin{equation}
\label{eq:rate}
\Gamma (\eta_h \rightarrow \gamma\gamma)= \pi\alpha^2Q_h^4M_{\eta_h}^3(F_{\eta_h}(0,0))^2 \, ,
\end{equation}
where $M_{\eta_h}$ is the meson mass and $Q_h$ the quark electric charge. 

As discussed in~\cite{Colquhoun:2023zbc}, a useful dimensionless ratio can be constructed from $F(0,0)$ and heavyonium decay constants and masses, that cancels the strong mass-dependence of its components and a lot of the QCD corrections and has a simple limit at leading-order in NRQCD: 
\begin{equation}
\label{eq:ratio}
R_{\eta_h} \equiv \frac{f_{\eta_h}}{F_{\eta_h}(0,0)M_{\eta_h}^2}= \frac{1}{2}(1+\mathcal{O}(\alpha_s)+\mathcal{O}(v^2/c^2)) \, .
\end{equation}
Here we calculate $R_{\eta_h}$ accurately in lattice QCD, both at the $b$ quark mass and for masses between $b$ and $c$, enabling us to plot out how the deviations from the leading-order result of 1/2 depend on heavy quark mass. We use the ratio at the $b$ and earlier HPQCD results for the $\eta_b$ decay constant~\cite{Hatton:2021dvg} to determine $\Gamma(\eta_b \rightarrow \gamma\gamma)$. 

\section{The lattice calculation} \label{sec:lattice}

We work on gluon field configurations that include $u$, $d$, $s$ and $c$ HISQ quarks in the sea generated by the MILC collaboration~\cite{Bazavov:2010ru,Bazavov:2012xda}, with parameters given in Table~\ref{tab:params}.  We calculate two- and three-point quark-line connected correlation functions by combining propagators for heavy valence quarks. The masses of these quarks are given in Table~\ref{tab:valence}; they are chosen to give a heavyonium pseudoscalar meson mass either that of the $\eta_b$ or 6.62 GeV or 4 GeV. 

\begin{table}
\caption{Parameters of MILC gluon configurations~\cite{Bazavov:2010ru,Bazavov:2012xda} 
with HISQ sea quark masses, $m_{l,s,c}^{\mathrm{sea}}$ ($m_u=m_d=m_l$) in lattice units.  
The lattice spacing, $a$, is given in units
of $w_0$~\cite{Borsanyi:2012zs} with
$w_0$=0.1715(9) fm from $f_{\pi}$~\cite{fkpi}. 
Sets  labelled `c' are `coarse' ($a \approx 0.12$ fm), `f' are `fine' ($a \approx 0.09$ fm), `sf' are
`superfine' ($a \approx 0.06$ fm), `uf' `ultrafine' ($a \approx 0.045$ fm) 
and `ef' `exafine' ($a \approx 0.03$ fm).  `5' or `phys' refer to the ratio $m_s^{\mathrm{sea}}/m_l^{\mathrm{sea}}$. 
$L_s$ and $L_t$  give the extent of the lattice in spatial and time directions. 
}
\begin{tabular}{lllllllll}
\hline
\hline
Set & $w_0/a$ & $am_{l}^{\mathrm{sea}}$ & $am_{s}^{\mathrm{sea}}$ & $am_{c}^{\mathrm{sea}}$ & $L_s$ & $L_t$ \\
\hline
 c-5 & 1.3826(11) & 0.0102 & 0.0509 & 0.635 & 24 & 64  \\
 c-phys & 1.4149(6) & 0.00184 & 0.0507 & 0.628 & 48 & 64 \\
 \hline
 f-5 & 1.9006(20) & 0.0074 & 0.037 & 0.440 & 32 & 96  \\
 f-phys & 1.9518(7) & 0.00120 & 0.0363 & 0.432 & 64 & 96 \\
\hline
 sf-5 & 2.8960(60) & 0.00480 & 0.0240 & 0.286 & 48  & 144 \\
 sf-phys & 3.0170(23) & 0.0008 & 0.022 & 0.260 & 96 & 192 \\
\hline
 uf-5 & 3.892(12) & 0.00316 & 0.0158 & 0.188 & 64 & 192 \\
\hline
 ef-5 & 5.243(16) & 0.00223 & 0.01115 & 0.1316 & 96 & 288 \\
\hline
\hline
\end{tabular}
\label{tab:params}
\end{table}

\begin{table}
\caption{Valence heavy quark masses in lattice units, for the three heavyonium pseudoscalar meson masses (for the `Goldstone' spin-taste) listed in the first row. The $\eta_b$ and 6.62 GeV cases are not accessible on the coarser sets since we take the limit $am_h < \pi/2$, following~\cite{Bazavov:2017lyh}. }
\begin{tabular}{llll}
\hline
\hline
  & & $am_h^{\mathrm{val}}$ \\
\hline
Set & $\eta_b$ \,\,\,\,& 6.62 GeV & 4.0 GeV \\
\hline
 c-5 &   & & 1.03 \\
 c-phys &  & & 1.0 \\
 \hline
 f-5 &  & 1.39 & 0.675 \\
 f-phys &  & 1.345 & 0.657\\
\hline
 sf-5 & 1.28 & 0.8 & 0.412  \\
 sf-phys & 1.21 & 0.76 & \\
\hline
 uf-5 & 0.88 & 0.545 & \\
\hline
 ef-5 &  0.622 & 0.403 & \\
\hline
\hline
\end{tabular}
\label{tab:valence}
\end{table}

\begin{figure}
		\includegraphics[width=0.4\textwidth]{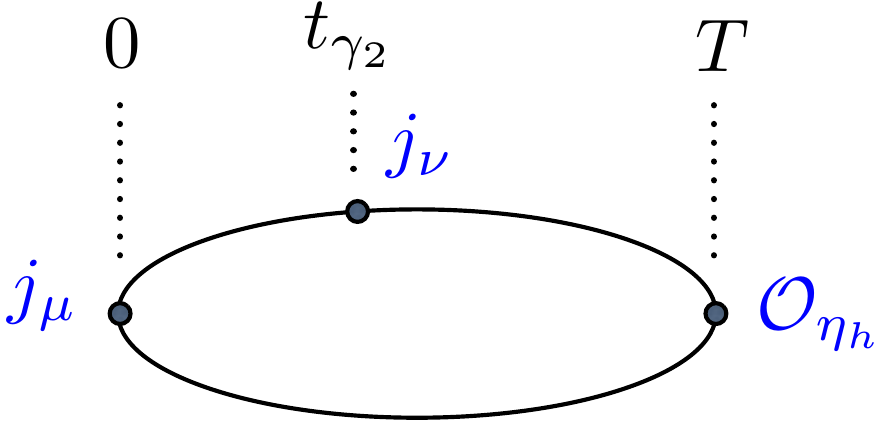}
		\caption{Schematic diagram of the connected 3-point correlation function between $O_{\eta_h}$ and two vector currents, see text. The lines between the operators represent $h$ quark propagators. We do not include any quark-line disconnected correlation functions in our calculation. }
		\label{fig:3pt-pic}
\end{figure}

We follow~\cite{Colquhoun:2023zbc} (Appendix B) in calculating the 3-point correlation function between a temporal axial vector current (coupling to the $\eta_h$) and two vector currents, shown schematically in Figure~\ref{fig:3pt-pic}. In spin-taste notation (see, for example, Appendix E of~\cite{Follana:2006rc}) we use operators $\gamma_5\gamma_t \otimes \gamma_x\gamma_z$, $\gamma_x\otimes \gamma_x$ and $\gamma_z\otimes\gamma_z$. One vector current is placed at the time origin, $0$, and the temporal axial current at $T$; 4--6 values of $T$ are used ranging from 0.7 fm upwards.  Propagators from $0$ and $T$ are combined with a vector current at $t_{\gamma_2}$, spatial momentum having been inserted in the propagator from 0 using twisted boundary conditions~\cite{Sachrajda:2004mi,Guadagnoli:2005be}. The spatial momentum is taken in the $y$-direction, orthogonal to both vector current polarisations, and its magnitude, $\omega$, is set equal to one-half the meson mass. The 3-point correlation function is then converted to a 2-point function between $0$ and $T$ by summing over all values of $t_{\gamma_2}$ from $-L_t/2+1$ to $+L_t/2-1$ with a weighting of $\exp(-\omega t_{\gamma_2})$. This sets photon 2 exactly on-shell~\cite{Ji:2001wha,Ji:2001nf,Dudek:2006ej}. For photon 1 $q_1^2=M^2_{\eta_h}(1-2\omega/M_{\eta_h})$, also zero up to tuning errors in $\omega$, which are at the level of our statistical errors here. The summand is strongly peaked around $t_{\gamma_2}=0$ and $T$ values are chosen to lie outside this peak~\cite{Colquhoun:2023zbc}. This gives a sizeable separation of the temporal axial current and the two vector currents and a good projection onto the ground-state $\eta_h$ meson. This is shown in the Supplementary materials~\cite{Supplementary} where a table of $T$ values is given. 

We fit this two-point function, $\tilde{C}(T)$, along with the full two-point function between two temporal axial currents, $C(t)$, to a sum of exponentials: 
\begin{equation}
\label{eq:fit}
C(t)=\sum_n a_n^2e^{-M_nt}\,\,;\,\,\tilde{C}(T)=\sum_n a_nb_n e^{-M_nT} ,
\end{equation}
where we have not shown the oscillating terms, included in our fit, that result from using staggered quarks. We use standard Bayesian fitting techniques for lattice QCD~\cite{Lepage:2001ym}. The ground-state amplitude, $b_0$, is then related to the matrix element for $\eta_h$ decay to two on-shell photons and we determine the form factor in lattice units as:
\begin{equation}
\label{eq:fit}
\frac{F(0,0)}{a}=b_0\sqrt{\frac{2}{aM_{\eta_h}}}\frac{L_s}{\theta\pi}Z_V^2\, .
\end{equation}
Here $\theta$ is the twist angle used and $Z_V$ is the renormalisation factor for the local vector current determined using an intermediate momentum-subtraction scheme known as RI-SMOM at a scale of 2 GeV (see Table 3 of ~\cite{Hatton:2019gha} and the Appendix of~\cite{Hatton:2020qhk}; we include the uncertainty in $Z_V$ in our calculation). By combining this with a simultaneous determination of the $\eta_h$ decay constant, $f_{\eta_h}$, from the annihilation amplitude of the `Goldstone' (spin-taste $\gamma_5\otimes\gamma_5$) $\eta_h$ meson, we can determine the ratio in Eq.~\eqref{eq:ratio}. Values for $R_{\eta_h}$ on each set and for each mass are given in the Supplementary materials~\cite{Supplementary}.

\section{Results and Conclusions} \label{sec:results}

\begin{figure}
		\includegraphics[width=0.4\textwidth]{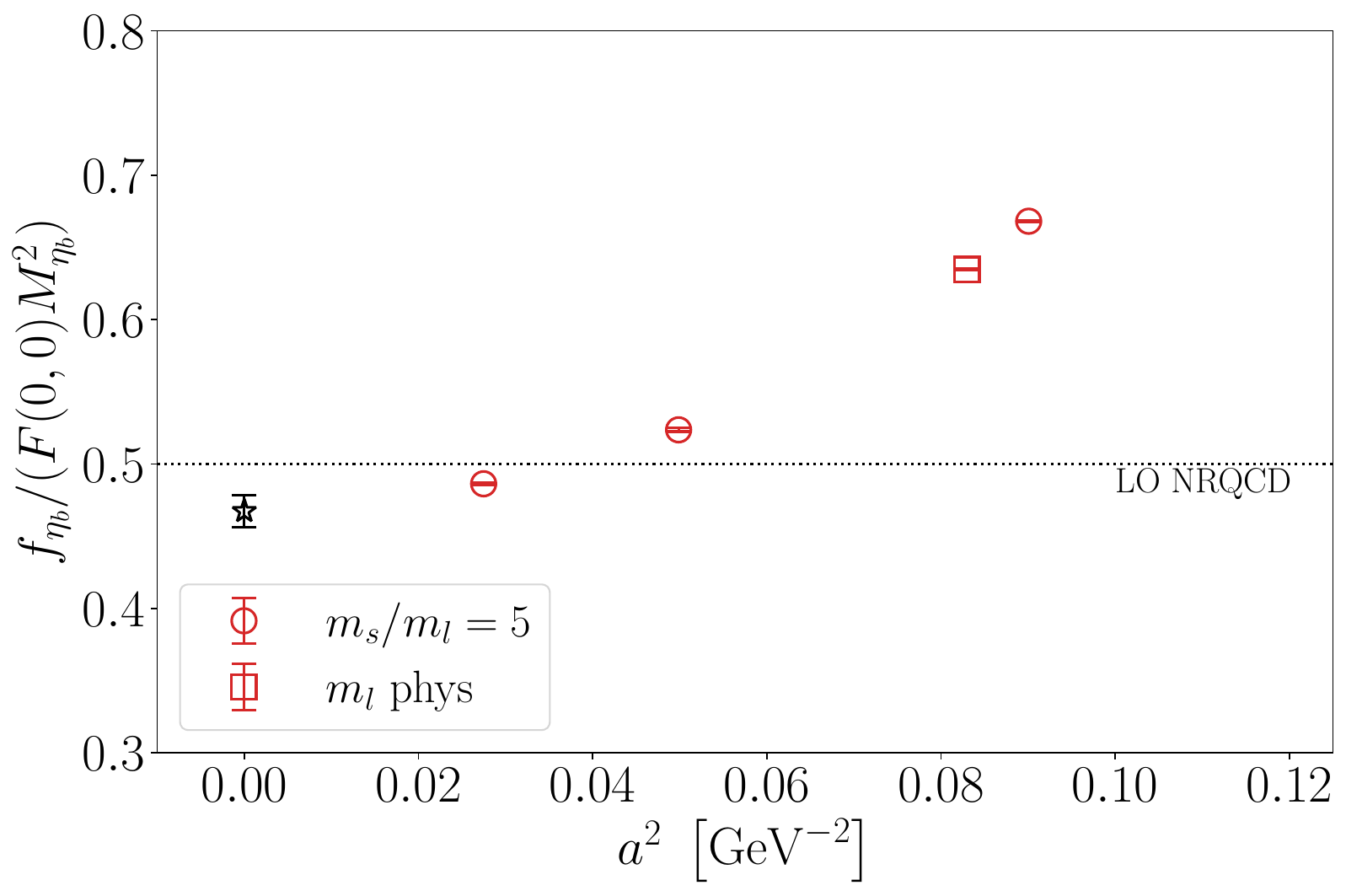}
		\caption{Lattice results (open red circles and square) and fit (red band) for the ratio $R_{\eta_h}$ (Eq.~\eqref{eq:ratio}) for the $\eta_b$. The black star and error bar denotes our result in the continuum limit. }
		\label{fig:fit}
\end{figure}

We fit the lattice results for $R_{\eta_h}$ for each meson mass as a function of $a$ and $m^{\mathrm{sea}}$ to determine a physical result in the continuum limit. The fit form, taken from~\cite{Colquhoun:2023zbc}, is:
\begin{align}
	&R_{\eta_h}^{\mathrm{latt}} = R_{\eta_h}^{\mathrm{phys}}{(1-q_1^2/M^2_{\mathrm{pole}})}\times \nonumber \\
	& \hspace{0mm} \Bigg[1+\sum_{i=1}^{i_{\text{max}}} \kappa_{a\Lambda}^{(i)}\left(a \Lambda\right)^{2 i} + \kappa_{\mathrm{val}, h} \delta^{\mathrm{val}, h} + \kappa_{\mathrm{sea}, c} \delta^{\mathrm{sea}, c} \nonumber \\
	& \hspace{3mm} + \kappa_{\mathrm{sea}, u d s}^{(0)} \delta^{\mathrm{sea}, u d s}\left\{1+\kappa_{\mathrm{sea}, u d s}^{(1)}(a\tilde{\Lambda} )^{2}+\kappa_{\mathrm{sea}, u d s}^{(2)}(a\tilde{\Lambda} )^{4}\right\}\Bigg] \label{eq:R_fit_form}
\end{align}
where the $\kappa_{a\Lambda}$ terms allow for discretisation effects, the $\kappa_h$ terms for mistuning of the heavy quark mass and the $\kappa_{\mathrm{sea}}$ terms for mistuning of the sea quark masses. We determine the scale, $\Lambda$, for the discretisation effects using the Empirical Bayes approach and find it is approximately equal to $M_{\eta_h}/\pi$, showing that it is set by the heavy quark mass, as expected. The number of discretisation terms, $i_{\mathrm{max}}$, is increased until the fit result, $R_{\eta_h}^{\mathrm{phys}}$, and its uncertainty is stable under the addition of further terms. We also test the addition of discretisation terms of the form $\alpha_s^n(a\Lambda)^2$ and find they have no impact. The factor $(1-q_1^2/M^2_{\mathrm{pole}})$ corrects for any small offshellness of photon 1 using a simple pole model. $M_{\mathrm{pole}}$ should be a heavyonium mass and we take it to be a fit parameter varying around $M_{\eta_h}$ with a width of 10\%. At the physical continuum limit $a=0$ and the quark masses are tuned to their physical values, taken from~\cite{Chakraborty:2014aca,McLean:2019sds,FermilabLattice:2018est}, so that all the $\delta=0$. Figure~\ref{fig:fit} shows our results at the $b$ quark mass and the continuum fit of Eq.~\eqref{eq:R_fit_form}. Further details on the fit form and stability tests are given in the Supplementary materials~\cite{Supplementary} along with plots of the fits for the other heavy quark mass cases.

\begin{figure}
		\includegraphics[width=0.45\textwidth]{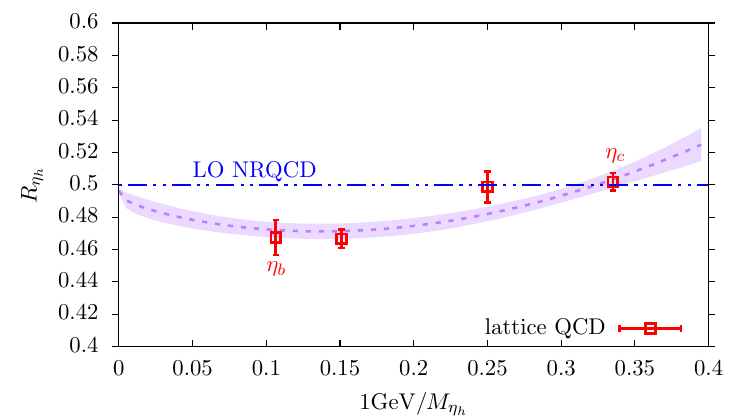}
		\caption{Lattice results (open red squares) for the ratio $R_{\eta_h}$ (Eq.~\eqref{eq:ratio}) for meson masses corresponding to those of $\eta_b$ and $\eta_c$ as well as two intermediate values of 4 GeV and 6.62 GeV, plotted against $1\,\mathrm{GeV}/M_{\eta_h}$. The $\eta_c$ result is derived from data in~\cite{Colquhoun:2023zbc}. The blue dash-dotted line shows the leading order NRQCD result of 0.5 and the purple dashed line and band is a simple fit to the lattice QCD results of NRQCD form adding in higher order corrections in $\alpha_s$ and powers of $1\,\mathrm{GeV}/M_{\eta_h}$ (see Eq.~\ref{eq:nrqcdfit}). }
		\label{fig:nrqcd}
\end{figure}

 The fit of Figure~\ref{fig:fit} allows us to obtain, at the physical point,
\begin{equation}
\label{eq:Rb}
R_{\eta_b}=  \finalRb\, \,; \,\, F(0,0)_{\eta_b} = \finalFb \, ,
\end{equation}
where we have used the value for the $\eta_b$ decay constant from lattice QCD~\cite{Hatton:2021dvg} of 724(12) MeV along with the experimental $\eta_b$ mass~\cite{ParticleDataGroup:2024cfk} to convert $R_{\eta_b}$ into the form factor. This enables us to predict a decay rate from Eq.~\eqref{eq:rate} of 
\begin{equation}
\label{eq:brate}
\finalWidthEtabGammaGamma \, .
\end{equation}
This is the first calculation of this quantity from lattice QCD. We use $1/\alpha=133.1$~\cite{ParticleDataGroup:2024cfk,Pivovarov:2000cr} in the $\overline{\mathrm{MS}}$ scheme, taking the scale of $\alpha$ as $0.26M_{\eta_b}$, following the Brodsky-Lepage-Mackenzie scale-setting procedure~\cite{Brodsky:1982gc}\footnote{This procedure uses the fermionic loop corrections to determine the momentum flowing in the leading-order diagram. The example given in~\cite{Brodsky:1982gc} is that of $\eta_c\to gg$ but the same result would be obtained for $\eta_c\to \gamma\gamma$ because the fermionic loops would appear in the same way.}. We include a systematic uncertainty of 0.2\% ($=\alpha/\pi$) from missing higher-order QED effects, noting that the lattice QCD+quenched QED calculation of~\cite{Hatton:2021dvg} found that giving the $b$ quarks electric charge had a less-than-0.1\% effect on $f_{\eta_b}$. The QED corrections to $F(0,0)$ and $f_{\eta_b}$ should be very similar given their relationship in $R_{\eta_h}$ (Eq.~\eqref{eq:ratio}), in which the QED corrections will largely cancel, as they do for QCD. We expect the uncertainty from missing quark-line disconnected diagrams to be negligible compared to this; the lattice QCD determination of the $\Upsilon-\eta_b$ mass splitting agrees within a few MeV (0.02\% of $M_{\eta_b}$) with experiment from connected correlators alone~\cite{Hatton:2021dvg}, showing that the impact of disconnected contributions on $\eta_b$ properties is very small (and less than for $\eta_c$~\cite{Hatton:2020qhk}). Two further systematic errors that we expect to be negligible are those from missing $b$ quarks in the sea\footnote{The impact of missing heavy sea quarks can be approximated by a change to the heavy quark potential proportional to $\delta^3(r)$ and with a coefficient $\sim$280 times smaller~\cite{Davies:2010ip} than the spin-dependent hyperfine potential of similar functional form. This induces a 7\% difference between the $\eta_b$ and $\Upsilon$ decay constants~\cite{Hatton:2021dvg}, giving an estimated 0.025\% effect on F(0,0) from missing sea $b$ quarks.} and the incomplete sampling of possible topological charge configurations on the finest lattices,  which should have little impact on heavyonium mesons~\cite{Bernard:2017npd}. Our total uncertainty on $\Gamma(\eta_b\rightarrow \gamma\gamma)$ is 6\%. 

The ratio of the $\Upsilon$ leptonic width, determined in lattice QCD in~\cite{Hatton:2021dvg}, to our result for the $\eta_b \to 2\gamma$ width in Eq.~\eqref{eq:brate} is $\finalupsratio$ (the result using the experimental $\Upsilon$ leptonic width~\cite{ParticleDataGroup:2024cfk,CLEO:2006uhx} is $\upsexptratio$). This is significantly less than the leading-order NRQCD result of 3~\cite{Czarnecki:2001zc}. The calculation of corrections to the leading-order value agree with our lattice QCD result but are much less accurate~\cite{ALEPH:2002xdz,Kiyo:2010jm}.

Figure~\ref{fig:nrqcd} shows our result for $R_{\eta_b}$ along with the $R_{\eta_h}$ results at meson masses of 6.62 GeV and 4 GeV. We also include a value for the $\eta_c$ case, calculating the ratio $R_{\eta_c}$ from lattice QCD results obtained for~\cite{Colquhoun:2023zbc} (that paper instead quotes results using the $J/\psi$ decay constant and mass). 
We see a mild dependence on heavy quark mass with $R_{\eta_h}$ increasing as $M_{\eta_h}$ falls. 

In the infinite quark mass (leading-order NRQCD) limit a value of 0.5 is expected with radiative and relativistic corrections at $b$ and $c$ masses. These can be sizeable, individually $\mathcal{O}(30\%)$ at next-to-leading order for the $\eta_c$, for example, but cancellation between different terms can reduce their net effect~\cite{Czarnecki:2001zc}. We see this happening in Figure~\ref{fig:nrqcd} with the $\eta_c$ result very close to leading-order NRQCD but the $\eta_b$ result below 0.5. The purple dashed line shows a simple uncorrelated fit to the lattice QCD results of form 
\begin{equation}
\label{eq:nrqcdfit}
R_{\eta_h}(M_{\eta_h}) =0.5\left(1 + \sum_{i=1,2}a_i\alpha_s^i+\sum_{j=1,2}b_j\left[\frac{1\,\mathrm{GeV}}{M_{\eta_h}}\right]^j\right)\, ,
\end{equation}
where $\alpha_s\equiv\overline{\alpha}_s(M_{\eta_h}/2)$.
From perturbative calculations in~\cite{Barbieri:1979be,Braaten:1995ej,Czarnecki:2001zc} we can derive $a_1=-0.099$. $a_2$ is given prior 0(1) and the $b_j$, 0(2). The fit has a $\chi^2/\mathrm{dof}=1.3/[4]$ and favours a small $1/M$, but a sizeable positive $1/M^2$, correction to leading-order NRQCD. We emphasise that, without the lattice QCD results, relative systematic uncertainties of at least $\pm 1/M_{\eta_h}$ would need to be taken on the leading-order NRQCD result and at the $\eta_b$ this gives a value in the range 0.45--0.55, much less accurate than that determined here.

By combining our result for $\Gamma(\eta_b \rightarrow \gamma\gamma)$ from Eq.~\eqref{eq:brate} with an NRQCD calculation of the branching fraction~\cite{Brambilla:2018tyu}  (using naive non-abelianisation resummation) updated to include color-octet matrix elements~\cite{Chung:2023bjr} we obtain a result for the $\eta_b$ total width of $\finalTotalWidthEtab$. This agrees with, but with 6\% total uncertainty is a lot more accurate than, the current experimental average value of $10^{+5}_{-4}\,\mathrm{MeV}$ for this quantity~\cite{ParticleDataGroup:2024cfk}.

{\it{Acknowledgements}}
We thank the MILC collaboration for making publicly available their gauge configurations and their code MILC-7.7.11 \cite{MILCgithub}. We are grateful to Hee Sok Chung for useful comments.
This work was performed using the Cambridge Service for Data Driven Discovery (CSD3), part of which is operated by the University of Cambridge Research Computing Service on behalf of the Science and Technology Facilities Council (STFC) DiRAC HPC Facility. The DiRAC component of CSD3 was funded by BEIS capital funding via STFC capital grants ST/P002307/1 and ST/R002452/1 and STFC operations grant ST/R00689X/1. DiRAC is part of the National e-Infrastructure. 
We are grateful to the CSD3 support staff for assistance.
Funding for this work came from STFC grant ST/T000945/1.

\bibliography{bib}{}

\begin{thebibliography}{37}%
\makeatletter
\providecommand \@ifxundefined [1]{%
 \@ifx{#1\undefined}
}%
\providecommand \@ifnum [1]{%
 \ifnum #1\expandafter \@firstoftwo
 \else \expandafter \@secondoftwo
 \fi
}%
\providecommand \@ifx [1]{%
 \ifx #1\expandafter \@firstoftwo
 \else \expandafter \@secondoftwo
 \fi
}%
\providecommand \natexlab [1]{#1}%
\providecommand \enquote  [1]{``#1''}%
\providecommand \bibnamefont  [1]{#1}%
\providecommand \bibfnamefont [1]{#1}%
\providecommand \citenamefont [1]{#1}%
\providecommand \href@noop [0]{\@secondoftwo}%
\providecommand \href [0]{\begingroup \@sanitize@url \@href}%
\providecommand \@href[1]{\@@startlink{#1}\@@href}%
\providecommand \@@href[1]{\endgroup#1\@@endlink}%
\providecommand \@sanitize@url [0]{\catcode `\\12\catcode `\$12\catcode
  `\&12\catcode `\#12\catcode `\^12\catcode `\_12\catcode `\%12\relax}%
\providecommand \@@startlink[1]{}%
\providecommand \@@endlink[0]{}%
\providecommand \url  [0]{\begingroup\@sanitize@url \@url }%
\providecommand \@url [1]{\endgroup\@href {#1}{\urlprefix }}%
\providecommand \urlprefix  [0]{URL }%
\providecommand \Eprint [0]{\href }%
\providecommand \doibase [0]{https://doi.org/}%
\providecommand \selectlanguage [0]{\@gobble}%
\providecommand \bibinfo  [0]{\@secondoftwo}%
\providecommand \bibfield  [0]{\@secondoftwo}%
\providecommand \translation [1]{[#1]}%
\providecommand \BibitemOpen [0]{}%
\providecommand \bibitemStop [0]{}%
\providecommand \bibitemNoStop [0]{.\EOS\space}%
\providecommand \EOS [0]{\spacefactor3000\relax}%
\providecommand \BibitemShut  [1]{\csname bibitem#1\endcsname}%
\let\auto@bib@innerbib\@empty
\bibitem [{\citenamefont {Colquhoun}\ \emph {et~al.}(2023)\citenamefont
  {Colquhoun}, \citenamefont {Cooper}, \citenamefont {Davies},\ and\
  \citenamefont {Lepage}}]{Colquhoun:2023zbc}%
  \BibitemOpen
  \bibfield  {author} {\bibinfo {author} {\bibfnamefont {B.}~\bibnamefont
  {Colquhoun}}, \bibinfo {author} {\bibfnamefont {L.~J.}\ \bibnamefont
  {Cooper}}, \bibinfo {author} {\bibfnamefont {C.~T.~H.}\ \bibnamefont
  {Davies}},\ and\ \bibinfo {author} {\bibfnamefont {G.~P.}\ \bibnamefont
  {Lepage}} (\bibinfo {collaboration} {HPQCD}),\ }\href
  {https://doi.org/10.1103/PhysRevD.108.014513} {\bibfield  {journal} {\bibinfo
   {journal} {Phys. Rev. D}\ }\textbf {\bibinfo {volume} {108}},\ \bibinfo
  {pages} {014513} (\bibinfo {year} {2023})},\ \Eprint
  {https://arxiv.org/abs/2305.06231} {arXiv:2305.06231 [hep-lat]} \BibitemShut
  {NoStop}%
\bibitem [{\citenamefont {Navas}\ \emph {et~al.}(2024)\citenamefont {Navas}
  \emph {et~al.}}]{ParticleDataGroup:2024cfk}%
  \BibitemOpen
  \bibfield  {author} {\bibinfo {author} {\bibfnamefont {S.}~\bibnamefont
  {Navas}} \emph {et~al.} (\bibinfo {collaboration} {Particle Data Group}),\
  }\href {https://doi.org/10.1103/PhysRevD.110.030001} {\bibfield  {journal}
  {\bibinfo  {journal} {Phys. Rev. D}\ }\textbf {\bibinfo {volume} {110}},\
  \bibinfo {pages} {030001} (\bibinfo {year} {2024})}\BibitemShut {NoStop}%
\bibitem [{\citenamefont {Heister}\ \emph {et~al.}(2002)\citenamefont {Heister}
  \emph {et~al.}}]{ALEPH:2002xdz}%
  \BibitemOpen
  \bibfield  {author} {\bibinfo {author} {\bibfnamefont {A.}~\bibnamefont
  {Heister}} \emph {et~al.} (\bibinfo {collaboration} {ALEPH}),\ }\href
  {https://doi.org/10.1016/S0370-2693(02)01329-1} {\bibfield  {journal}
  {\bibinfo  {journal} {Phys. Lett. B}\ }\textbf {\bibinfo {volume} {530}},\
  \bibinfo {pages} {56} (\bibinfo {year} {2002})},\ \Eprint
  {https://arxiv.org/abs/hep-ex/0202011} {arXiv:hep-ex/0202011} \BibitemShut
  {NoStop}%
\bibitem [{\citenamefont {Lange}\ and\ \citenamefont
  {Prencipe}(2022)}]{Belletalk}%
  \BibitemOpen
  \bibfield  {author} {\bibinfo {author} {\bibfnamefont {J.~S.}\ \bibnamefont
  {Lange}}\ and\ \bibinfo {author} {\bibfnamefont {E.}~\bibnamefont
  {Prencipe}},\ }\href@noop {} {\bibfield  {journal} {\bibinfo  {journal}
  {Hadron Spectroscopy: The next big steps, workshop at Mainz Institute for
  Theoretical Physics}\ } (\bibinfo {year} {2022})}\BibitemShut {NoStop}%
\bibitem [{\citenamefont {Follana}\ \emph {et~al.}(2007)\citenamefont
  {Follana}, \citenamefont {Mason}, \citenamefont {Davies}, \citenamefont
  {Hornbostel}, \citenamefont {Lepage}, \citenamefont {Shigemitsu},
  \citenamefont {Trottier},\ and\ \citenamefont {Wong}}]{Follana:2006rc}%
  \BibitemOpen
  \bibfield  {author} {\bibinfo {author} {\bibfnamefont {E.}~\bibnamefont
  {Follana}}, \bibinfo {author} {\bibfnamefont {Q.}~\bibnamefont {Mason}},
  \bibinfo {author} {\bibfnamefont {C.}~\bibnamefont {Davies}}, \bibinfo
  {author} {\bibfnamefont {K.}~\bibnamefont {Hornbostel}}, \bibinfo {author}
  {\bibfnamefont {G.}~\bibnamefont {Lepage}}, \bibinfo {author} {\bibfnamefont
  {J.}~\bibnamefont {Shigemitsu}}, \bibinfo {author} {\bibfnamefont
  {H.}~\bibnamefont {Trottier}},\ and\ \bibinfo {author} {\bibfnamefont
  {K.}~\bibnamefont {Wong}} (\bibinfo {collaboration} {HPQCD, UKQCD}),\ }\href
  {https://doi.org/10.1103/PhysRevD.75.054502} {\bibfield  {journal} {\bibinfo
  {journal} {Phys. Rev. D}\ }\textbf {\bibinfo {volume} {75}},\ \bibinfo
  {pages} {054502} (\bibinfo {year} {2007})},\ \Eprint
  {https://arxiv.org/abs/hep-lat/0610092} {arXiv:hep-lat/0610092} \BibitemShut
  {NoStop}%
\bibitem [{\citenamefont {Monahan}\ \emph {et~al.}(2013)\citenamefont
  {Monahan}, \citenamefont {Shigemitsu},\ and\ \citenamefont
  {Horgan}}]{Monahan:2012dq}%
  \BibitemOpen
  \bibfield  {author} {\bibinfo {author} {\bibfnamefont {C.}~\bibnamefont
  {Monahan}}, \bibinfo {author} {\bibfnamefont {J.}~\bibnamefont
  {Shigemitsu}},\ and\ \bibinfo {author} {\bibfnamefont {R.}~\bibnamefont
  {Horgan}},\ }\href {https://doi.org/10.1103/PhysRevD.87.034017} {\bibfield
  {journal} {\bibinfo  {journal} {Phys. Rev. D}\ }\textbf {\bibinfo {volume}
  {87}},\ \bibinfo {pages} {034017} (\bibinfo {year} {2013})},\ \Eprint
  {https://arxiv.org/abs/1211.6966} {arXiv:1211.6966 [hep-lat]} \BibitemShut
  {NoStop}%
\bibitem [{\citenamefont {Colquhoun}\ \emph {et~al.}(2025)\citenamefont
  {Colquhoun}, \citenamefont {Davies}, \citenamefont {Hatton},\ and\
  \citenamefont {Lepage}}]{Colquhoun:2025pnh}%
  \BibitemOpen
  \bibfield  {author} {\bibinfo {author} {\bibfnamefont {B.}~\bibnamefont
  {Colquhoun}}, \bibinfo {author} {\bibfnamefont {C.~T.~H.}\ \bibnamefont
  {Davies}}, \bibinfo {author} {\bibfnamefont {D.}~\bibnamefont {Hatton}},\
  and\ \bibinfo {author} {\bibfnamefont {G.~P.}\ \bibnamefont {Lepage}},\
  }\href@noop {} {\  (\bibinfo {year} {2025})},\ \Eprint
  {https://arxiv.org/abs/2508.02862} {arXiv:2508.02862 [hep-lat]} \BibitemShut
  {NoStop}%
\bibitem [{\citenamefont {Hatton}\ \emph {et~al.}(2021)\citenamefont {Hatton},
  \citenamefont {Davies}, \citenamefont {Koponen}, \citenamefont {Lepage},\
  and\ \citenamefont {Lytle}}]{Hatton:2021dvg}%
  \BibitemOpen
  \bibfield  {author} {\bibinfo {author} {\bibfnamefont {D.}~\bibnamefont
  {Hatton}}, \bibinfo {author} {\bibfnamefont {C.~T.~H.}\ \bibnamefont
  {Davies}}, \bibinfo {author} {\bibfnamefont {J.}~\bibnamefont {Koponen}},
  \bibinfo {author} {\bibfnamefont {G.~P.}\ \bibnamefont {Lepage}},\ and\
  \bibinfo {author} {\bibfnamefont {A.~T.}\ \bibnamefont {Lytle}} (\bibinfo
  {collaboration} {HPQCD}),\ }\href
  {https://doi.org/10.1103/PhysRevD.103.054512} {\bibfield  {journal} {\bibinfo
   {journal} {Phys. Rev. D}\ }\textbf {\bibinfo {volume} {103}},\ \bibinfo
  {pages} {054512} (\bibinfo {year} {2021})},\ \Eprint
  {https://arxiv.org/abs/2101.08103} {arXiv:2101.08103 [hep-lat]} \BibitemShut
  {NoStop}%
\bibitem [{\citenamefont {Bazavov}\ \emph {et~al.}(2010)\citenamefont {Bazavov}
  \emph {et~al.}}]{Bazavov:2010ru}%
  \BibitemOpen
  \bibfield  {author} {\bibinfo {author} {\bibfnamefont {A.}~\bibnamefont
  {Bazavov}} \emph {et~al.} (\bibinfo {collaboration} {MILC}),\ }\href
  {https://doi.org/10.1103/PhysRevD.82.074501} {\bibfield  {journal} {\bibinfo
  {journal} {Phys. Rev.}\ }\textbf {\bibinfo {volume} {D82}},\ \bibinfo {pages}
  {074501} (\bibinfo {year} {2010})},\ \Eprint
  {https://arxiv.org/abs/1004.0342} {arXiv:1004.0342 [hep-lat]} \BibitemShut
  {NoStop}%
\bibitem [{\citenamefont {Bazavov}\ \emph {et~al.}(2013)\citenamefont {Bazavov}
  \emph {et~al.}}]{Bazavov:2012xda}%
  \BibitemOpen
  \bibfield  {author} {\bibinfo {author} {\bibfnamefont {A.}~\bibnamefont
  {Bazavov}} \emph {et~al.} (\bibinfo {collaboration} {MILC}),\ }\href
  {https://doi.org/10.1103/PhysRevD.87.054505} {\bibfield  {journal} {\bibinfo
  {journal} {Phys. Rev.}\ }\textbf {\bibinfo {volume} {D87}},\ \bibinfo {pages}
  {054505} (\bibinfo {year} {2013})},\ \Eprint
  {https://arxiv.org/abs/1212.4768} {arXiv:1212.4768 [hep-lat]} \BibitemShut
  {NoStop}%
\bibitem [{\citenamefont {Bors\'anyi}\ \emph {et~al.}(2012)\citenamefont
  {Bors\'anyi} \emph {et~al.}}]{Borsanyi:2012zs}%
  \BibitemOpen
  \bibfield  {author} {\bibinfo {author} {\bibfnamefont {S.}~\bibnamefont
  {Bors\'anyi}} \emph {et~al.},\ }\href
  {https://doi.org/10.1007/JHEP09(2012)010} {\bibfield  {journal} {\bibinfo
  {journal} {JHEP}\ }\textbf {\bibinfo {volume} {09}},\ \bibinfo {pages}
  {010}},\ \Eprint {https://arxiv.org/abs/1203.4469} {arXiv:1203.4469
  [hep-lat]} \BibitemShut {NoStop}%
\bibitem [{\citenamefont {Dowdall}\ \emph {et~al.}(2013)\citenamefont
  {Dowdall}, \citenamefont {Davies}, \citenamefont {Lepage},\ and\
  \citenamefont {McNeile}}]{fkpi}%
  \BibitemOpen
  \bibfield  {author} {\bibinfo {author} {\bibfnamefont {R.}~\bibnamefont
  {Dowdall}}, \bibinfo {author} {\bibfnamefont {C.}~\bibnamefont {Davies}},
  \bibinfo {author} {\bibfnamefont {G.}~\bibnamefont {Lepage}},\ and\ \bibinfo
  {author} {\bibfnamefont {C.}~\bibnamefont {McNeile}} (\bibinfo
  {collaboration} {HPQCD}),\ }\href
  {https://doi.org/10.1103/PhysRevD.88.074504} {\bibfield  {journal} {\bibinfo
  {journal} {Phys. Rev. D}\ }\textbf {\bibinfo {volume} {88}},\ \bibinfo
  {pages} {074504} (\bibinfo {year} {2013})},\ \Eprint
  {https://arxiv.org/abs/1303.1670} {arXiv:1303.1670 [hep-lat]} \BibitemShut
  {NoStop}%
\bibitem [{\citenamefont {Bazavov}\ \emph
  {et~al.}(2018{\natexlab{a}})\citenamefont {Bazavov} \emph
  {et~al.}}]{Bazavov:2017lyh}%
  \BibitemOpen
  \bibfield  {author} {\bibinfo {author} {\bibfnamefont {A.}~\bibnamefont
  {Bazavov}} \emph {et~al.} (\bibinfo {collaboration} {Fermilab Lattice,
  MILC}),\ }\href {https://doi.org/10.1103/PhysRevD.98.074512} {\bibfield
  {journal} {\bibinfo  {journal} {Phys. Rev. D}\ }\textbf {\bibinfo {volume}
  {98}},\ \bibinfo {pages} {074512} (\bibinfo {year} {2018}{\natexlab{a}})},\
  \Eprint {https://arxiv.org/abs/1712.09262} {arXiv:1712.09262 [hep-lat]}
  \BibitemShut {NoStop}%
\bibitem [{\citenamefont {Sachrajda}\ and\ \citenamefont
  {Villadoro}(2005)}]{Sachrajda:2004mi}%
  \BibitemOpen
  \bibfield  {author} {\bibinfo {author} {\bibfnamefont {C.}~\bibnamefont
  {Sachrajda}}\ and\ \bibinfo {author} {\bibfnamefont {G.}~\bibnamefont
  {Villadoro}},\ }\href {https://doi.org/10.1016/j.physletb.2005.01.033}
  {\bibfield  {journal} {\bibinfo  {journal} {Phys. Lett. B}\ }\textbf
  {\bibinfo {volume} {609}},\ \bibinfo {pages} {73} (\bibinfo {year} {2005})},\
  \Eprint {https://arxiv.org/abs/hep-lat/0411033} {arXiv:hep-lat/0411033}
  \BibitemShut {NoStop}%
\bibitem [{\citenamefont {Guadagnoli}\ \emph {et~al.}(2006)\citenamefont
  {Guadagnoli}, \citenamefont {Mescia},\ and\ \citenamefont
  {Simula}}]{Guadagnoli:2005be}%
  \BibitemOpen
  \bibfield  {author} {\bibinfo {author} {\bibfnamefont {D.}~\bibnamefont
  {Guadagnoli}}, \bibinfo {author} {\bibfnamefont {F.}~\bibnamefont {Mescia}},\
  and\ \bibinfo {author} {\bibfnamefont {S.}~\bibnamefont {Simula}},\ }\href
  {https://doi.org/10.1103/PhysRevD.73.114504} {\bibfield  {journal} {\bibinfo
  {journal} {Phys. Rev. D}\ }\textbf {\bibinfo {volume} {73}},\ \bibinfo
  {pages} {114504} (\bibinfo {year} {2006})},\ \Eprint
  {https://arxiv.org/abs/hep-lat/0512020} {arXiv:hep-lat/0512020} \BibitemShut
  {NoStop}%
\bibitem [{\citenamefont {Ji}\ and\ \citenamefont
  {Jung}(2001{\natexlab{a}})}]{Ji:2001wha}%
  \BibitemOpen
  \bibfield  {author} {\bibinfo {author} {\bibfnamefont {X.}~\bibnamefont
  {Ji}}\ and\ \bibinfo {author} {\bibfnamefont {C.}~\bibnamefont {Jung}},\
  }\href {https://doi.org/10.1103/PhysRevLett.86.208} {\bibfield  {journal}
  {\bibinfo  {journal} {Phys. Rev. Lett.}\ }\textbf {\bibinfo {volume} {86}},\
  \bibinfo {pages} {208} (\bibinfo {year} {2001}{\natexlab{a}})},\ \Eprint
  {https://arxiv.org/abs/hep-lat/0101014} {arXiv:hep-lat/0101014} \BibitemShut
  {NoStop}%
\bibitem [{\citenamefont {Ji}\ and\ \citenamefont
  {Jung}(2001{\natexlab{b}})}]{Ji:2001nf}%
  \BibitemOpen
  \bibfield  {author} {\bibinfo {author} {\bibfnamefont {X.}~\bibnamefont
  {Ji}}\ and\ \bibinfo {author} {\bibfnamefont {C.}~\bibnamefont {Jung}},\
  }\href {https://doi.org/10.1103/PhysRevD.64.034506} {\bibfield  {journal}
  {\bibinfo  {journal} {Phys. Rev. D}\ }\textbf {\bibinfo {volume} {64}},\
  \bibinfo {pages} {034506} (\bibinfo {year} {2001}{\natexlab{b}})},\ \Eprint
  {https://arxiv.org/abs/hep-lat/0103007} {arXiv:hep-lat/0103007} \BibitemShut
  {NoStop}%
\bibitem [{\citenamefont {Dudek}\ \emph {et~al.}(2006)\citenamefont {Dudek},
  \citenamefont {Edwards},\ and\ \citenamefont {Richards}}]{Dudek:2006ej}%
  \BibitemOpen
  \bibfield  {author} {\bibinfo {author} {\bibfnamefont {J.~J.}\ \bibnamefont
  {Dudek}}, \bibinfo {author} {\bibfnamefont {R.~G.}\ \bibnamefont {Edwards}},\
  and\ \bibinfo {author} {\bibfnamefont {D.~G.}\ \bibnamefont {Richards}},\
  }\href {https://doi.org/10.1103/PhysRevD.73.074507} {\bibfield  {journal}
  {\bibinfo  {journal} {Phys. Rev. D}\ }\textbf {\bibinfo {volume} {73}},\
  \bibinfo {pages} {074507} (\bibinfo {year} {2006})},\ \Eprint
  {https://arxiv.org/abs/hep-ph/0601137} {arXiv:hep-ph/0601137} \BibitemShut
  {NoStop}%
\bibitem [{Sup()}]{Supplementary}%
  \BibitemOpen
  \href@noop {} {}\bibinfo {note} {See Supplementary Material at
  https://arxiv.org/abs/2410.24041 for tabulated data and further details of
  the fits and analysis.}\BibitemShut {Stop}%
\bibitem [{\citenamefont {Lepage}\ \emph {et~al.}(2002)\citenamefont {Lepage},
  \citenamefont {Clark}, \citenamefont {Davies}, \citenamefont {Hornbostel},
  \citenamefont {Mackenzie}, \citenamefont {Morningstar},\ and\ \citenamefont
  {Trottier}}]{Lepage:2001ym}%
  \BibitemOpen
  \bibfield  {author} {\bibinfo {author} {\bibfnamefont {G.~P.}\ \bibnamefont
  {Lepage}}, \bibinfo {author} {\bibfnamefont {B.}~\bibnamefont {Clark}},
  \bibinfo {author} {\bibfnamefont {C.~T.~H.}\ \bibnamefont {Davies}}, \bibinfo
  {author} {\bibfnamefont {K.}~\bibnamefont {Hornbostel}}, \bibinfo {author}
  {\bibfnamefont {P.~B.}\ \bibnamefont {Mackenzie}}, \bibinfo {author}
  {\bibfnamefont {C.}~\bibnamefont {Morningstar}},\ and\ \bibinfo {author}
  {\bibfnamefont {H.}~\bibnamefont {Trottier}} (\bibinfo {collaboration}
  {HPQCD}),\ }\href {https://doi.org/10.1016/S0920-5632(01)01638-3} {\bibfield
  {journal} {\bibinfo  {journal} {Nucl. Phys. B Proc. Suppl.}\ }\textbf
  {\bibinfo {volume} {106}},\ \bibinfo {pages} {12} (\bibinfo {year} {2002})},\
  \Eprint {https://arxiv.org/abs/hep-lat/0110175} {arXiv:hep-lat/0110175}
  \BibitemShut {NoStop}%
\bibitem [{\citenamefont {Hatton}\ \emph {et~al.}(2019)\citenamefont {Hatton},
  \citenamefont {Davies}, \citenamefont {Lepage},\ and\ \citenamefont
  {Lytle}}]{Hatton:2019gha}%
  \BibitemOpen
  \bibfield  {author} {\bibinfo {author} {\bibfnamefont {D.}~\bibnamefont
  {Hatton}}, \bibinfo {author} {\bibfnamefont {C.~T.~H.}\ \bibnamefont
  {Davies}}, \bibinfo {author} {\bibfnamefont {G.~P.}\ \bibnamefont {Lepage}},\
  and\ \bibinfo {author} {\bibfnamefont {A.~T.}\ \bibnamefont {Lytle}}
  (\bibinfo {collaboration} {HPQCD}),\ }\href
  {https://doi.org/10.1103/PhysRevD.100.114513} {\bibfield  {journal} {\bibinfo
   {journal} {Phys. Rev. D}\ }\textbf {\bibinfo {volume} {100}},\ \bibinfo
  {pages} {114513} (\bibinfo {year} {2019})},\ \Eprint
  {https://arxiv.org/abs/1909.00756} {arXiv:1909.00756 [hep-lat]} \BibitemShut
  {NoStop}%
\bibitem [{\citenamefont {Hatton}\ \emph {et~al.}(2020)\citenamefont {Hatton},
  \citenamefont {Davies}, \citenamefont {Galloway}, \citenamefont {Koponen},
  \citenamefont {Lepage},\ and\ \citenamefont {Lytle}}]{Hatton:2020qhk}%
  \BibitemOpen
  \bibfield  {author} {\bibinfo {author} {\bibfnamefont {D.}~\bibnamefont
  {Hatton}}, \bibinfo {author} {\bibfnamefont {C.}~\bibnamefont {Davies}},
  \bibinfo {author} {\bibfnamefont {B.}~\bibnamefont {Galloway}}, \bibinfo
  {author} {\bibfnamefont {J.}~\bibnamefont {Koponen}}, \bibinfo {author}
  {\bibfnamefont {G.}~\bibnamefont {Lepage}},\ and\ \bibinfo {author}
  {\bibfnamefont {A.}~\bibnamefont {Lytle}} (\bibinfo {collaboration}
  {HPQCD}),\ }\href {https://doi.org/10.1103/PhysRevD.102.054511} {\bibfield
  {journal} {\bibinfo  {journal} {Phys. Rev. D}\ }\textbf {\bibinfo {volume}
  {102}},\ \bibinfo {pages} {054511} (\bibinfo {year} {2020})},\ \Eprint
  {https://arxiv.org/abs/2005.01845} {arXiv:2005.01845 [hep-lat]} \BibitemShut
  {NoStop}%
\bibitem [{\citenamefont {Chakraborty}\ \emph {et~al.}(2015)\citenamefont
  {Chakraborty}, \citenamefont {Davies}, \citenamefont {Galloway},
  \citenamefont {Knecht}, \citenamefont {Koponen}, \citenamefont {Donald},
  \citenamefont {Dowdall}, \citenamefont {Lepage},\ and\ \citenamefont
  {McNeile}}]{Chakraborty:2014aca}%
  \BibitemOpen
  \bibfield  {author} {\bibinfo {author} {\bibfnamefont {B.}~\bibnamefont
  {Chakraborty}}, \bibinfo {author} {\bibfnamefont {C.~T.~H.}\ \bibnamefont
  {Davies}}, \bibinfo {author} {\bibfnamefont {B.}~\bibnamefont {Galloway}},
  \bibinfo {author} {\bibfnamefont {P.}~\bibnamefont {Knecht}}, \bibinfo
  {author} {\bibfnamefont {J.}~\bibnamefont {Koponen}}, \bibinfo {author}
  {\bibfnamefont {G.~C.}\ \bibnamefont {Donald}}, \bibinfo {author}
  {\bibfnamefont {R.~J.}\ \bibnamefont {Dowdall}}, \bibinfo {author}
  {\bibfnamefont {G.~P.}\ \bibnamefont {Lepage}},\ and\ \bibinfo {author}
  {\bibfnamefont {C.}~\bibnamefont {McNeile}} (\bibinfo {collaboration}
  {HPQCD}),\ }\href {https://doi.org/10.1103/PhysRevD.91.054508} {\bibfield
  {journal} {\bibinfo  {journal} {Phys. Rev. D}\ }\textbf {\bibinfo {volume}
  {91}},\ \bibinfo {pages} {054508} (\bibinfo {year} {2015})},\ \Eprint
  {https://arxiv.org/abs/1408.4169} {arXiv:1408.4169 [hep-lat]} \BibitemShut
  {NoStop}%
\bibitem [{\citenamefont {McLean}\ \emph {et~al.}(2019)\citenamefont {McLean},
  \citenamefont {Davies}, \citenamefont {Lytle},\ and\ \citenamefont
  {Koponen}}]{McLean:2019sds}%
  \BibitemOpen
  \bibfield  {author} {\bibinfo {author} {\bibfnamefont {E.}~\bibnamefont
  {McLean}}, \bibinfo {author} {\bibfnamefont {C.~T.~H.}\ \bibnamefont
  {Davies}}, \bibinfo {author} {\bibfnamefont {A.~T.}\ \bibnamefont {Lytle}},\
  and\ \bibinfo {author} {\bibfnamefont {J.}~\bibnamefont {Koponen}} (\bibinfo
  {collaboration} {HPQCD}),\ }\href
  {https://doi.org/10.1103/PhysRevD.99.114512} {\bibfield  {journal} {\bibinfo
  {journal} {Phys. Rev. D}\ }\textbf {\bibinfo {volume} {99}},\ \bibinfo
  {pages} {114512} (\bibinfo {year} {2019})},\ \Eprint
  {https://arxiv.org/abs/1904.02046} {arXiv:1904.02046 [hep-lat]} \BibitemShut
  {NoStop}%
\bibitem [{\citenamefont {Bazavov}\ \emph
  {et~al.}(2018{\natexlab{b}})\citenamefont {Bazavov} \emph
  {et~al.}}]{FermilabLattice:2018est}%
  \BibitemOpen
  \bibfield  {author} {\bibinfo {author} {\bibfnamefont {A.}~\bibnamefont
  {Bazavov}} \emph {et~al.} (\bibinfo {collaboration} {Fermilab Lattice, MILC,
  TUMQCD}),\ }\href {https://doi.org/10.1103/PhysRevD.98.054517} {\bibfield
  {journal} {\bibinfo  {journal} {Phys. Rev. D}\ }\textbf {\bibinfo {volume}
  {98}},\ \bibinfo {pages} {054517} (\bibinfo {year} {2018}{\natexlab{b}})},\
  \Eprint {https://arxiv.org/abs/1802.04248} {arXiv:1802.04248 [hep-lat]}
  \BibitemShut {NoStop}%
\bibitem [{\citenamefont {Pivovarov}(2002)}]{Pivovarov:2000cr}%
  \BibitemOpen
  \bibfield  {author} {\bibinfo {author} {\bibfnamefont {A.~A.}\ \bibnamefont
  {Pivovarov}},\ }\href {https://doi.org/10.1134/1.1495645} {\bibfield
  {journal} {\bibinfo  {journal} {Phys. Atom. Nucl.}\ }\textbf {\bibinfo
  {volume} {65}},\ \bibinfo {pages} {1319} (\bibinfo {year} {2002})},\ \Eprint
  {https://arxiv.org/abs/hep-ph/0011135} {arXiv:hep-ph/0011135} \BibitemShut
  {NoStop}%
\bibitem [{\citenamefont {Brodsky}\ \emph {et~al.}(1983)\citenamefont
  {Brodsky}, \citenamefont {Lepage},\ and\ \citenamefont
  {Mackenzie}}]{Brodsky:1982gc}%
  \BibitemOpen
  \bibfield  {author} {\bibinfo {author} {\bibfnamefont {S.~J.}\ \bibnamefont
  {Brodsky}}, \bibinfo {author} {\bibfnamefont {G.~P.}\ \bibnamefont
  {Lepage}},\ and\ \bibinfo {author} {\bibfnamefont {P.~B.}\ \bibnamefont
  {Mackenzie}},\ }\href {https://doi.org/10.1103/PhysRevD.28.228} {\bibfield
  {journal} {\bibinfo  {journal} {Phys. Rev. D}\ }\textbf {\bibinfo {volume}
  {28}},\ \bibinfo {pages} {228} (\bibinfo {year} {1983})}\BibitemShut
  {NoStop}%
\bibitem [{\citenamefont {Davies}\ \emph {et~al.}(2010)\citenamefont {Davies},
  \citenamefont {McNeile}, \citenamefont {Follana}, \citenamefont {Lepage},
  \citenamefont {Na},\ and\ \citenamefont {Shigemitsu}}]{Davies:2010ip}%
  \BibitemOpen
  \bibfield  {author} {\bibinfo {author} {\bibfnamefont {C.~T.~H.}\
  \bibnamefont {Davies}}, \bibinfo {author} {\bibfnamefont {C.}~\bibnamefont
  {McNeile}}, \bibinfo {author} {\bibfnamefont {E.}~\bibnamefont {Follana}},
  \bibinfo {author} {\bibfnamefont {G.~P.}\ \bibnamefont {Lepage}}, \bibinfo
  {author} {\bibfnamefont {H.}~\bibnamefont {Na}},\ and\ \bibinfo {author}
  {\bibfnamefont {J.}~\bibnamefont {Shigemitsu}},\ }\href
  {https://doi.org/10.1103/PhysRevD.82.114504} {\bibfield  {journal} {\bibinfo
  {journal} {Phys. Rev. D}\ }\textbf {\bibinfo {volume} {82}},\ \bibinfo
  {pages} {114504} (\bibinfo {year} {2010})},\ \Eprint
  {https://arxiv.org/abs/1008.4018} {arXiv:1008.4018 [hep-lat]} \BibitemShut
  {NoStop}%
\bibitem [{\citenamefont {Bernard}\ and\ \citenamefont
  {Toussaint}(2018)}]{Bernard:2017npd}%
  \BibitemOpen
  \bibfield  {author} {\bibinfo {author} {\bibfnamefont {C.}~\bibnamefont
  {Bernard}}\ and\ \bibinfo {author} {\bibfnamefont {D.}~\bibnamefont
  {Toussaint}} (\bibinfo {collaboration} {MILC}),\ }\href
  {https://doi.org/10.1103/PhysRevD.97.074502} {\bibfield  {journal} {\bibinfo
  {journal} {Phys. Rev. D}\ }\textbf {\bibinfo {volume} {97}},\ \bibinfo
  {pages} {074502} (\bibinfo {year} {2018})},\ \Eprint
  {https://arxiv.org/abs/1707.05430} {arXiv:1707.05430 [hep-lat]} \BibitemShut
  {NoStop}%
\bibitem [{\citenamefont {Besson}\ \emph {et~al.}(2007)\citenamefont {Besson}
  \emph {et~al.}}]{CLEO:2006uhx}%
  \BibitemOpen
  \bibfield  {author} {\bibinfo {author} {\bibfnamefont {D.}~\bibnamefont
  {Besson}} \emph {et~al.} (\bibinfo {collaboration} {CLEO}),\ }\href
  {https://doi.org/10.1103/PhysRevLett.98.052002} {\bibfield  {journal}
  {\bibinfo  {journal} {Phys. Rev. Lett.}\ }\textbf {\bibinfo {volume} {98}},\
  \bibinfo {pages} {052002} (\bibinfo {year} {2007})},\ \Eprint
  {https://arxiv.org/abs/hep-ex/0607019} {arXiv:hep-ex/0607019} \BibitemShut
  {NoStop}%
\bibitem [{\citenamefont {Czarnecki}\ and\ \citenamefont
  {Melnikov}(2001)}]{Czarnecki:2001zc}%
  \BibitemOpen
  \bibfield  {author} {\bibinfo {author} {\bibfnamefont {A.}~\bibnamefont
  {Czarnecki}}\ and\ \bibinfo {author} {\bibfnamefont {K.}~\bibnamefont
  {Melnikov}},\ }\href {https://doi.org/10.1016/S0370-2693(01)01129-7}
  {\bibfield  {journal} {\bibinfo  {journal} {Phys. Lett. B}\ }\textbf
  {\bibinfo {volume} {519}},\ \bibinfo {pages} {212} (\bibinfo {year}
  {2001})},\ \Eprint {https://arxiv.org/abs/hep-ph/0109054}
  {arXiv:hep-ph/0109054} \BibitemShut {NoStop}%
\bibitem [{\citenamefont {Kiyo}\ \emph {et~al.}(2010)\citenamefont {Kiyo},
  \citenamefont {Pineda},\ and\ \citenamefont {Signer}}]{Kiyo:2010jm}%
  \BibitemOpen
  \bibfield  {author} {\bibinfo {author} {\bibfnamefont {Y.}~\bibnamefont
  {Kiyo}}, \bibinfo {author} {\bibfnamefont {A.}~\bibnamefont {Pineda}},\ and\
  \bibinfo {author} {\bibfnamefont {A.}~\bibnamefont {Signer}},\ }\href
  {https://doi.org/10.1016/j.nuclphysb.2010.08.007} {\bibfield  {journal}
  {\bibinfo  {journal} {Nucl. Phys. B}\ }\textbf {\bibinfo {volume} {841}},\
  \bibinfo {pages} {231} (\bibinfo {year} {2010})},\ \Eprint
  {https://arxiv.org/abs/1006.2685} {arXiv:1006.2685 [hep-ph]} \BibitemShut
  {NoStop}%
\bibitem [{\citenamefont {Barbieri}\ \emph {et~al.}(1979)\citenamefont
  {Barbieri}, \citenamefont {d'Emilio}, \citenamefont {Curci},\ and\
  \citenamefont {Remiddi}}]{Barbieri:1979be}%
  \BibitemOpen
  \bibfield  {author} {\bibinfo {author} {\bibfnamefont {R.}~\bibnamefont
  {Barbieri}}, \bibinfo {author} {\bibfnamefont {E.}~\bibnamefont {d'Emilio}},
  \bibinfo {author} {\bibfnamefont {G.}~\bibnamefont {Curci}},\ and\ \bibinfo
  {author} {\bibfnamefont {E.}~\bibnamefont {Remiddi}},\ }\href
  {https://doi.org/10.1016/0550-3213(79)90047-6} {\bibfield  {journal}
  {\bibinfo  {journal} {Nucl. Phys. B}\ }\textbf {\bibinfo {volume} {154}},\
  \bibinfo {pages} {535} (\bibinfo {year} {1979})}\BibitemShut {NoStop}%
\bibitem [{\citenamefont {Braaten}\ and\ \citenamefont
  {Fleming}(1995)}]{Braaten:1995ej}%
  \BibitemOpen
  \bibfield  {author} {\bibinfo {author} {\bibfnamefont {E.}~\bibnamefont
  {Braaten}}\ and\ \bibinfo {author} {\bibfnamefont {S.}~\bibnamefont
  {Fleming}},\ }\href {https://doi.org/10.1103/PhysRevD.52.181} {\bibfield
  {journal} {\bibinfo  {journal} {Phys. Rev. D}\ }\textbf {\bibinfo {volume}
  {52}},\ \bibinfo {pages} {181} (\bibinfo {year} {1995})},\ \Eprint
  {https://arxiv.org/abs/hep-ph/9501296} {arXiv:hep-ph/9501296} \BibitemShut
  {NoStop}%
\bibitem [{\citenamefont {Brambilla}\ \emph {et~al.}(2018)\citenamefont
  {Brambilla}, \citenamefont {Chung},\ and\ \citenamefont
  {Komijani}}]{Brambilla:2018tyu}%
  \BibitemOpen
  \bibfield  {author} {\bibinfo {author} {\bibfnamefont {N.}~\bibnamefont
  {Brambilla}}, \bibinfo {author} {\bibfnamefont {H.~S.}\ \bibnamefont
  {Chung}},\ and\ \bibinfo {author} {\bibfnamefont {J.}~\bibnamefont
  {Komijani}},\ }\href {https://doi.org/10.1103/PhysRevD.98.114020} {\bibfield
  {journal} {\bibinfo  {journal} {Phys. Rev. D}\ }\textbf {\bibinfo {volume}
  {98}},\ \bibinfo {pages} {114020} (\bibinfo {year} {2018})},\ \Eprint
  {https://arxiv.org/abs/1810.02586} {arXiv:1810.02586 [hep-ph]} \BibitemShut
  {NoStop}%
\bibitem [{\citenamefont {Chung}(2024)}]{Chung:2023bjr}%
  \BibitemOpen
  \bibfield  {author} {\bibinfo {author} {\bibfnamefont {H.~S.}\ \bibnamefont
  {Chung}},\ }\href {https://doi.org/10.1103/PhysRevD.109.054001} {\bibfield
  {journal} {\bibinfo  {journal} {Phys. Rev. D}\ }\textbf {\bibinfo {volume}
  {109}},\ \bibinfo {pages} {054001} (\bibinfo {year} {2024})},\ \Eprint
  {https://arxiv.org/abs/2312.10601} {arXiv:2312.10601 [hep-ph]} \BibitemShut
  {NoStop}%
\bibitem [{MIL()}]{MILCgithub}%
  \BibitemOpen
  \href@noop {} {}\bibinfo {note} {MILC Code Repository,
  https://github.com/milc-qcd}\BibitemShut {NoStop}%
\end{thebibliography}%
\bibliographystyle{apsrev4-2}

\end{document}